# Large Polaron Generation and Dynamics in 3D Metal-Halide Perovskites

Walter P.D. Wong,[1,†] Jun Yin,[2,†] Bhumika Chaudhary,[3,4] Chin Xin Yu,[4] Daniele Cortecchia,[3,4] Shu-Zee A. Lo,[5] Andrew C. Grimsdale,[1] Guglielmo Lanzani,[4,6,7] and Cesare Soci[3,4,5*]

[1] School of Materials Science and Engineering, Nanyang Technological University, 50 Nanyang Avenue, Singapore 639798

[2] Division of Physical Science and Engineering, King Abdullah University of Science and Technology, Thuwal 23955-6900, Saudi Arabia

[3] Interdisciplinary Graduate School, Nanyang Technological University, 50 Nanyang Avenue, Singapore 639798

[4] Energy Research Institute (ERI@N), Research Techno Plaza, X-Frontier Block, Level 5, 50 Nanyang Drive, Singapore 6375553

[5] Division of Physics and Applied Physics, School of Physical and Mathematical Sciences, Nanyang Technological University, 21 Nanyang Link, Singapore, 637371

[6] Dipartimento di Fisica, Politecnico di Milano, Piazza Leonardo da Vinci 32, Milano, Italy

[7] CNST@PoliMi, Istituto Italiano di Tecnologia (IIT), Via Giovanni Pascoli 70/3, Milano, Italy

[†]These authors contributed equally

*Corresponding author: csoci@ntu.edu.sg

**Abstract**

In recent years, metal halide perovskites have generated tremendous interest for optoelectronic applications and their underlying fundamental properties. Due to the large electron-phonon coupling characteristic of soft lattices, self-trapping phenomena are expected to dominate hybrid perovskite photoexcitation dynamics. Yet, while the photogeneration of small polarons was proven in low dimensional perovskites, the nature of polaron excitations in technologically relevant 3D perovskites, and their influence on charge carrier transport, remain elusive. In this study, we used a combination of first principle calculations and advanced spectroscopy techniques spanning the entire optical frequency range to pin down polaron features in 3D metal halide perovskites. Mid-infrared photoinduced absorption shows the photogeneration of states associated to low energy intragap electronic transitions with lifetime up to the ms time scale, and vibrational mode renormalization in both frequency and amplitude. Density functional theory supports the assignment of the spectroscopic features to large polarons leading to new intra gap transitions, hardening of phonon mode frequency, and renormalization of the oscillator strength. Theory provides quantitative estimates of the charge carrier masses and mobilities increase upon polaron formation, confirming experimental results. Overall, this work contributes to complete the scenario of elementary photoexcitations in metal halide perovskites and highlights the importance of polaronic transport in perovskite-based optoelectronic devices.

Metal halide perovskites (MHP) are attracting enormous interest as solution processed[1-2] active materials in a variety of device applications.[3-15] The prototype compound is methylammonium lead iodide (MAPbI3), a three-dimensional (3D) hybrid perovskite known for its remarkable transport properties such as long charge carrier lifetime >1 µs[16] and diffusion length >3 µm,[17-19] small non-radiative bimolecular recombination coefficient $\gamma$ of the order of $10^{-10}$ $cm^3s^{-1}$, comparable to those of crystalline semiconductors such as GaAs,[20-21] and apparent insensitivity to defects.[22-25] However, the nature of the charge carriers responsible for such remarkable performance is still under debate. Reported charge carrier mobilities in perovskite films between $10^{-4}$ and 10 $cm^2V^{-1}s^{-1}$, depending on measurement technique[16-17, 20, 26-30] are indeed in net contrast with the prediction of the Langevin model for free charge carrier recombination in semiconductors, where $\gamma/\mu = e/(\varepsilon_0\varepsilon_r)$.[27] The origin of such apparent inconsistency is still debated,[31] and points toward the role of polaronic protection of charge carriers in the perovskite lattice.[32-33]

Due to the highly deformable and polar nature of the metal halide framework, MHPs are prone to lattice relaxation, which is expected to cause self-trapping of the elementary excitations into phonon-dressed localised states.[34-36] In low dimensional perovskites (e.g. 2D and 1D structures), where Coulomb interactions are enhanced by reduced dielectric screening and quantum confinement effects, the formation of self-trapped excitons (*small polarons*) manifests itself in apparent radiative recombination effects, such as white-light emission in compounds like (EDBE)PbBr4.[35, 37-39] Here, *ab initio* calculations have indicated that the excess charge is spatially confined to one crystal unit cell or less, inducing local distortions of the lead-halide framework. Photoexcitation in 2D perovskites gives rise to photoinduced lattice deformations[34-35, 40] associated to polaron exciton states, with a characteristic fine structure in the absorption line-shapes.[41] On the other hand, for charge carriers in 3D perovskites,[42] due to the long range electron phonon interaction typical of ionic crystals, polarons may extend over several lattice sites (*large polarons*). Large polarons can display band-like coherent transport with mobility >1 cm$^2$/Vs, that falls with increasing temperature.[33, 43] In addition, the electrostatic screening brought about by the ionic (polar) lattice deformation could hinder Coulomb mediated processes such as Auger cooling and Langevin recombination.[44]

Several recent works have postulated the existence of large polaron states in 3D MHPs. Among these are the electrical conductivity exhibiting characteristic power-law temperature dependence,[45] the steady-state photoinduced absorption (PIA) in visible[46] and far infrared (FIR) spectral range,[47] Terahertz spectroscopic investigations,[48-49] and time-resolved optical Kerr effect spectroscopy (TR-OKE)[50] and optical absorption spectroscopy (TR-OAS).[51] Formation of long-lived energetic carriers with ~10$^2$ ps lifetime in MAPbBr$_3$ and FAPbBr$_3$ points towards the screening of charge carriers in form of large polarons through liquid-like collective reorientation of the molecular dipoles.[52] Impulsive Raman scattering performed on MAPbBr$_3$[53] revealed the presence of coherent phonon modes generated via displacive excitation that is indicative of strong electron-phonon coupling, instrumental to polaron formation but not a proof of their existence. Similarly, a polaron state has been invoked to explain the octahedral distortion of PbI$_6^{4-}$ sustained by coherent vibrations of the Pb-I normal modes in MAPbI$_3$.[54] Nonetheless, the impact of carrier induced lattice distortion on polaron formation and charge transport in optoelectronic devices has remained elusive.[16]

In this work, we identify the geometrical pattern associated to large self-trapped polaron states in MAPbI$_3$ by combining an advanced theoretical investigation of charge-induced lattice relaxation with the experimental study of specific vibrational modes. We assign the unique spectroscopic signatures of polaron-induced lattice relaxation in the infrared spectral region and polaron-induced electronic transitions in the near-infrared to visible range. Based on this assignment, we further study polaron generation and long decay dynamics with femtosecond temporal resolution and present the first "real time" fast photocurrent transients, which directly measure charge carrier relaxation at the picosecond time scale and beyond. The correspondence between transient absorption and photocurrent decay kinetics supports the conjecture that polarons are primary charge carriers in 3D perovskites.

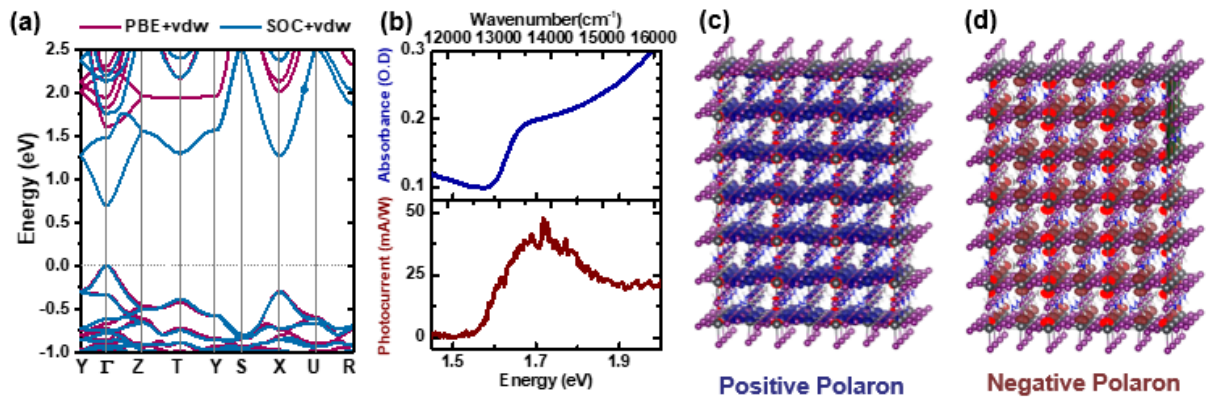

**Figure 1**. Electronic structure, absorption and photocurrent spectra of orthorhombic MAPbI$_3$. (a) Calculated band structures at GGA/PBE+vdw level without and with SOC effects; (b) linear absorption (top) and normalised steady-state photocurrent spectra obtained with illumination intensity of ~20 µW/cm$^2$ under applied field of 9 kV/cm (bottom); (c-d) electronic charge densities of positive (c) and negative (d) polaronic states for a 3×3×3 supercell calculated at the GGA/PBE level.

The band structure of orthorhombic-phase MAPbI$_3$ was calculated at GGA/PBE+vdW level without and with spin-orbit coupling (SOC) effects (Figure 1a). The resulting bandgap (1.61 eV, without SOC) is in good agreement with the experimental value deduced from the steady-state absorption spectra and the onset of steady-state photocurrent (Figure 1b). Inclusion of SOC in the band structure calculations considerably reduces the bandgap to 0.70 eV due to band splitting. As shown in Table 1, the extracted values of free carriers effective masses ($m^*$) along the M→Γ/Γ→Z directions are 0.166/0.253 (hole) and 0.121/0.204 (electron) for the low temperature orthorhombic-phase MAPbI$_3$. The similar values obtained for hole and electron effective masses confirm the ambipolar nature of charge transport in MAPbI$_3$, while the different values for different crystallographic directions reflects the intrinsic anisotropy of the 3D perovskite lattice. Similar results are obtained for the high temperature tetragonal phase, except for the inverted transport anisotropy.

**Table 1.** Calculated effective mass ($m^*$, ×$m_0$) and mobility (cm$^2$/V·s, average of different crystallographic directions) of free charges and polarons in tetragonal- and orthorhombic-phase MAPbI$_3$ (including spin-orbit coupling, effects).

| Type | Tetragonal | | | Orthorhombic | | |
|---|---|---|---|---|---|---|
| | Effective Mass | | Mobility (cm$^2$/V·s) | Effective Mass | | Mobility (cm$^2$/V·s) |
| | M(Y)→Γ | Γ→Z | | M(Y)→Γ | Γ→Z | |
| **Free Hole** | 0.240 | 0.180 | 332.2 | 0.166 | 0.253 | 1491.6 |
| **Positive Polaron** | 1.411 | 1.058 | 215.4 | 0.976 | 1.487 | 497.9 |
| **Free Electron** | 0.220 | 0.107 | 630.5 | 0.121 | 0.204 | 5122.0 |
| **Negative Polaron** | 1.293 | 0.629 | 225.9 | 0.711 | 1.199 | 550.7 |

Adding a positive/negative charge to the neutral lattice induces spontaneous structural relaxation of the supercell to a new equilibrium geometry, leading to the formation of positive and negative polarons (Figs. 1c and 1d). The new half-filled energy level associated to the negative polaron lies 0.187 eV below the conduction band minimum (CBM) while that associated to the positive polaron lies 0.235 eV above the valence band maximum (VBM). The resulting charge density distributions suggest that the self-trapped polaron states are localized in spatially distinct regions within ~2-3 inorganic atomic layers. Charge localization with comparable spatial extent was also observed in 3D and 2D perovskites when varying the Pb-I distances[35] or increasing the tilting of Pb-I-Pb angles.[55] Accordingly, the calculated polaronic effective masses in MAPbI$_3$ system are about 5 times larger than those of free carriers and calculated polaron mobilities are 1.5 to 8 times smaller than free charge carrier mobilities.

The electronic and vibrational structure of semiconductors are strongly affected by polaron formation: i. half-filled electronic states appear within the gap, giving rise to new optical transitions; ii. vibrational mode frequencies are renormalized by the change in bond geometry, including modes of the inorganic lattice and of the organic moiety that "feel" a deformed environment (*spectator modes*),[52, 54, 56] but are not directly coupled to the carrier; iii. geometric distortion induces local symmetry breaking that relaxes vibrational selection rules and eventually activates vibrational modes that are silent for bare states. To test all these specific predictions, we measured the infrared photoinduced absorption of $MAPbI_3$ over a broad spectral range (Fig. 2a).

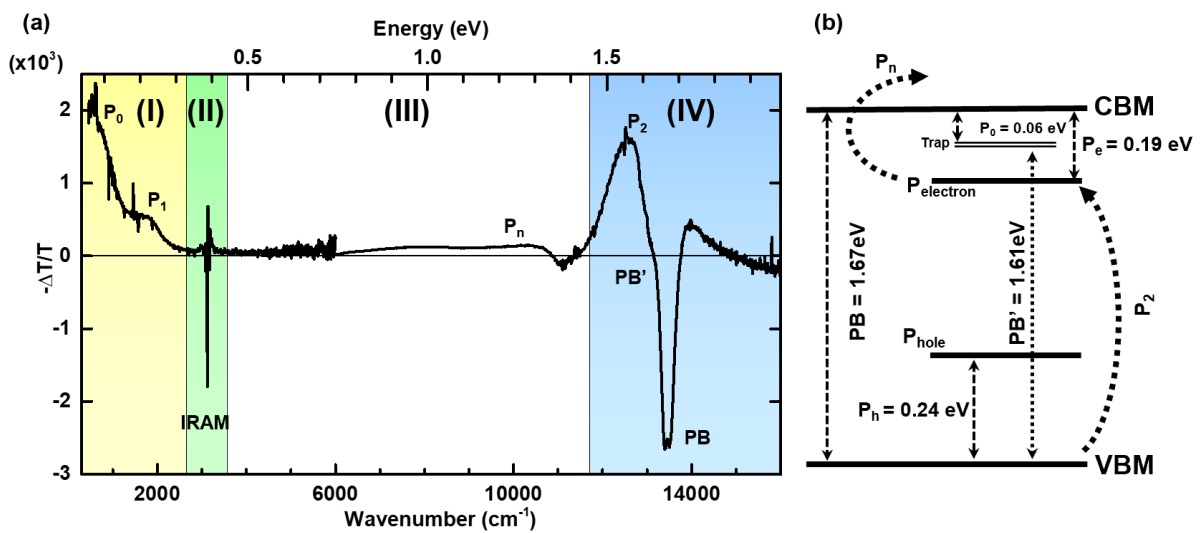

**Figure 2.** (a) cw-PIA spectrum of $MAPbI_3$. Region I (500–2600 $cm^{-1}$) shows the intragap polaron electronic transition $P_1=0.20$ eV and the trap state absorption $P_0=0.06$ eV, convoluted with infrared active vibrational (IRAV) modes. Region II (2600–3500 $cm^{-1}$) shows infrared active modulation (IRAM) of the C–H and N–H vibrational mode frequencies. Region III (3500–11700 $cm^{-1}$) indicates the presence of PIA transitions to higher lying states. Region IV (11700–16000 $cm^{-1}$) contains high energy intragap polaron transitions and photobleaching (PB) of ground state absorption. (b) Proposed electronic energy diagram and associated transitions, based on the calculated polaron state energies. $P_1$ is assigned to the convolution of positive polaron $P_h=0.24$ eV and negative polaron $P_e=0.187$ eV transitions above the valance band maximum (VBM) and below the conduction band minimum (CBM), respectively.

Low-temperature (T~80 K) steady-state continuous wave photoinduced absorption (cw-PIA) measurements of the MAPbI$_3$ film (see *Experimental Methods* section for details) show clear evidence for the generation of long-lived photoexcited states (with typical lifetimes of the order of milliseconds). Four regions of interest are identified in the cw-PIA spectrum in Fig. 2: a photoinduced absorption region between 500–2500 cm$^{-1}$ containing two absorption peaks (region I), a vibrational mode region at 3000-3300 cm$^{-1}$ (region II), a second photoinduced absorption region containing a broad (P$_n$) transition around 6000–11500 cm$^{-1}$ (region III), and a high energy modulation region just below the band gap, featuring photoinduced absorption (PA) and photobleaching (PB) (region IV). The concomitant disappearing of P$_1$, P$_2$ and P$_n$ at room temperature indicates their common origin (see Figure S3).

Figure 3a show details of the PIA spectrum in region I, featuring two photoinduced absorption bands centered around 516 cm$^{-1}$ (P$_0$) and 1650 cm$^1$ (P$_1$), as determined by Gaussian fitting curves in red, and a series of small features at 620 cm$^{-1}$, 908 cm$^{-1}$, 1254 cm$^{-1}$, 1460 cm$^{-1}$ and 1570 cm$^{-1}$ (solid black lines running across Fig 3a-c and d-f) that correspond to the amplitude modulation of the main IR active (IRAV) vibrational modes.[55] An exception is the photoinduced mode at 600 cm$^{-1}$ (within the range of C-N vibrational modes of the organic moiety) which corresponds to an *IR inactive* Raman mode; this suggests breaking of the inversion symmetry that governs mutually exclusive Raman and IR mode selection rules due to the local distortion of the lattice in the presence of the self-trapped charge.[57]

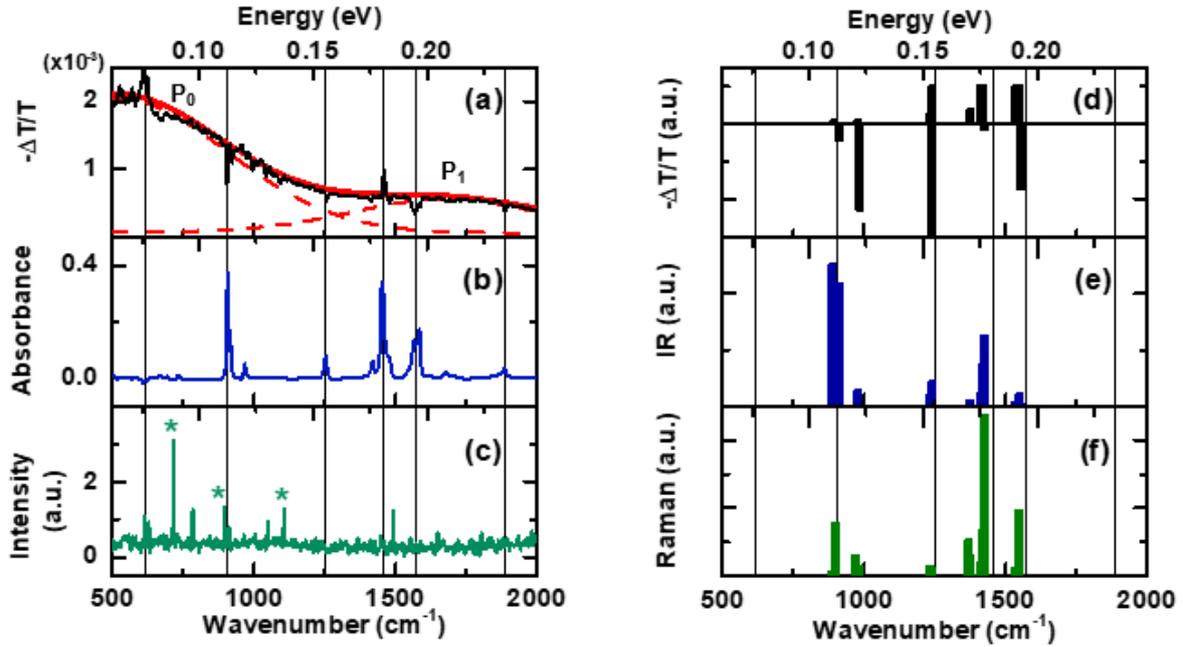

**Figure 3.** Experimental and simulated infrared active vibrational (IRAV) and Raman modes. (a) PIA spectrum (T=78 K) fitted to two Gaussian peaks centred around 1650 cm$^{-1}$ and 516 cm$^{-1}$ (red dashed lines), (b) FTIR spectrum (T=78 K) and (c) Raman spectrum ($\lambda_{exc}$=1024 nm, T=300 K), where green asterisks denote peaks from the CaF$_2$ substrate. (d) Simulated photoinduced absorption modes obtained as $-(\Delta T/T) = -(IR_{neut} - IR_{exc})/IR_{neut}$, where $IR_{exc}$ and $IR_{neut}$ are the IR mode intensities of excited and ground states, respectively; (e, f) calculated IR and Raman mode intensity spectra of orthorhombic MAPbI$_3$. The main vibrational modes are indicated by black vertical lines, showing the correspondence between IRAV modes and Raman modes which become IR active under photoexcitation.

DFT calculations accurately account for the experimental observations in region I of the spectrum. The broad PA band P$_1$ at 1650 cm$^{-1}$ (0.204 eV) corresponds to a convolution of electronic transitions from negative and positive polaron states at P$_e$=0.187 eV below the CBM and P$_h$=0.235 eV above the VBM, respectively. Conversely, the P$_0$ peak centred at 516 cm$^{-1}$ (0.06 eV), that does not emerge from the calculations, is likely due to a trap state.[25] Figs. 3e and 3f show the calculated IR and Raman spectra, which correspond to the experimental modes assigned to the organic moiety. The photoinduced modulation of the IRAV modes from 750 cm$^{-1}$ to 2000 cm$^{-1}$ is reproduced reasonably well by the calculations. Fig. 3d shows the spectrum obtained by computing the difference between the spectra calculated in the undistorted (Fig.

3e) and distorted perovskite lattices. The modulation sign is in agreement with the experiment for each mode, accounting for amplitude increase or decrease of the vibrational transitions upon polaron formation. Notably, this suggests that negative PA dips are due to a decrease of the vibrational transition oscillator strength, rather than to Fano resonances due to quantum interference of electronic and vibrational transitions.[58]

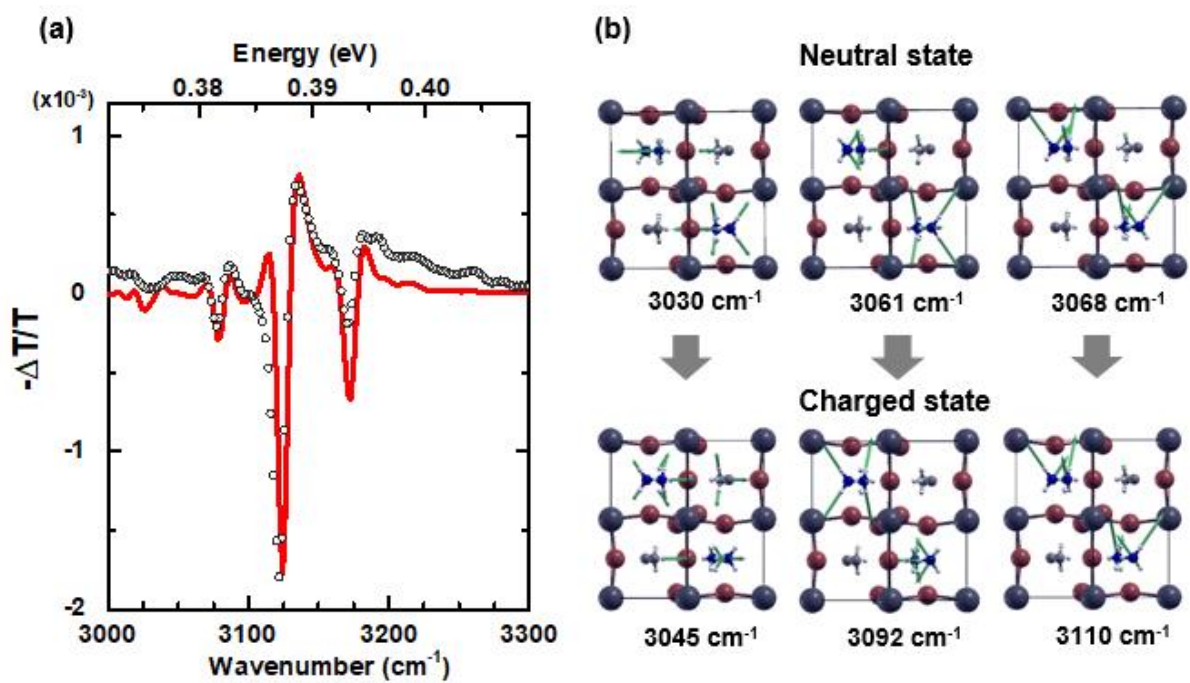

**Figure 4.** Infrared active modulation (IRAM) of C–H and N–H vibrational modes. (a) Experimental data fitted to a linear combination of first and second derivative of the IR absorption spectrum $f(\bar{v})$, $f_{der}(\bar{v}) = -0.149 f'(\bar{v}) + 0.314 f''(\bar{v})$. The first derivative indicates a spectral shift of the absorption peak, while the second derivative indicates its spectral broadening; in this case, C–H and N–H mode hardening upon photoexcitation induces a blue shift of the vibrational modes. (b) Representative IR vibrational frequencies and displacement vectors obtained from phonon calculations, and corresponding blue-shift of IR active mode frequencies in the charged state.

The orthorhombic-phase MAPbI$_3$ shows several IR peaks in the spectral region of 3000 - 3300 cm$^{-1}$, which can be assigned to both C–H and N–H stretching modes of the organic cations. The photoinduced modulation of these modes in region II is particularly interesting (Fig. 4). We fitted the experimental PA data in this region to its first and second derivative (Fig. 4a). The first derivative component of the absorption spectrum gives an indication of the spectral blue shift of the peak, while the second derivative component is related to the peak broadening[59]. Both effects can be attributed to the modulation of C-H and N-H bonds induced by the polaron which, by locally distorting the perovskite lattice, indirectly affect their vibrational potential through the hydrogen bonds between hydrogen atoms of -$NH_3$ and iodine atoms in charged $MAPbI_3$ (*spectator modes*).  Note that thermal modulation cannot account for this effect as it determines an opposite red shift of the mode frequency (mode softening), see Supplementary Material. In spite of the slight mismatch of calculated and experimental IR mode frequencies in both neutral and excited states (Figure 4b), the blue shift induced by the lattice distortion on the vibrational modes is correctly reproduced by the calculations. Thus, the observed long-lived vibrational frequency renormalization upon photoexcitation is a clear fingerprint of the generation of a geometrically relaxed state.

The concomitant photoinduced electronic spectrum, with intragap transitions, corroborate the assignment to photogenetated polarons. Details of the cw-PIA spectrum in region IV are shown in the top panel of Figure 5a. The spectrum can be fitted by three Gaussian curves, labelled $P_2$ (centered at ~1.60 eV), PB' (centered at ~1.61 eV), and PB (centered at ~1.67 eV), where $P_2$ corresponds well to the energetics determined for ground state to polaron transitions, while the PB and PB' peaks are assigned to valence to conduction band and valence to trap-state transitions, respectively (refer to the energy diagram in Fig. 2b).

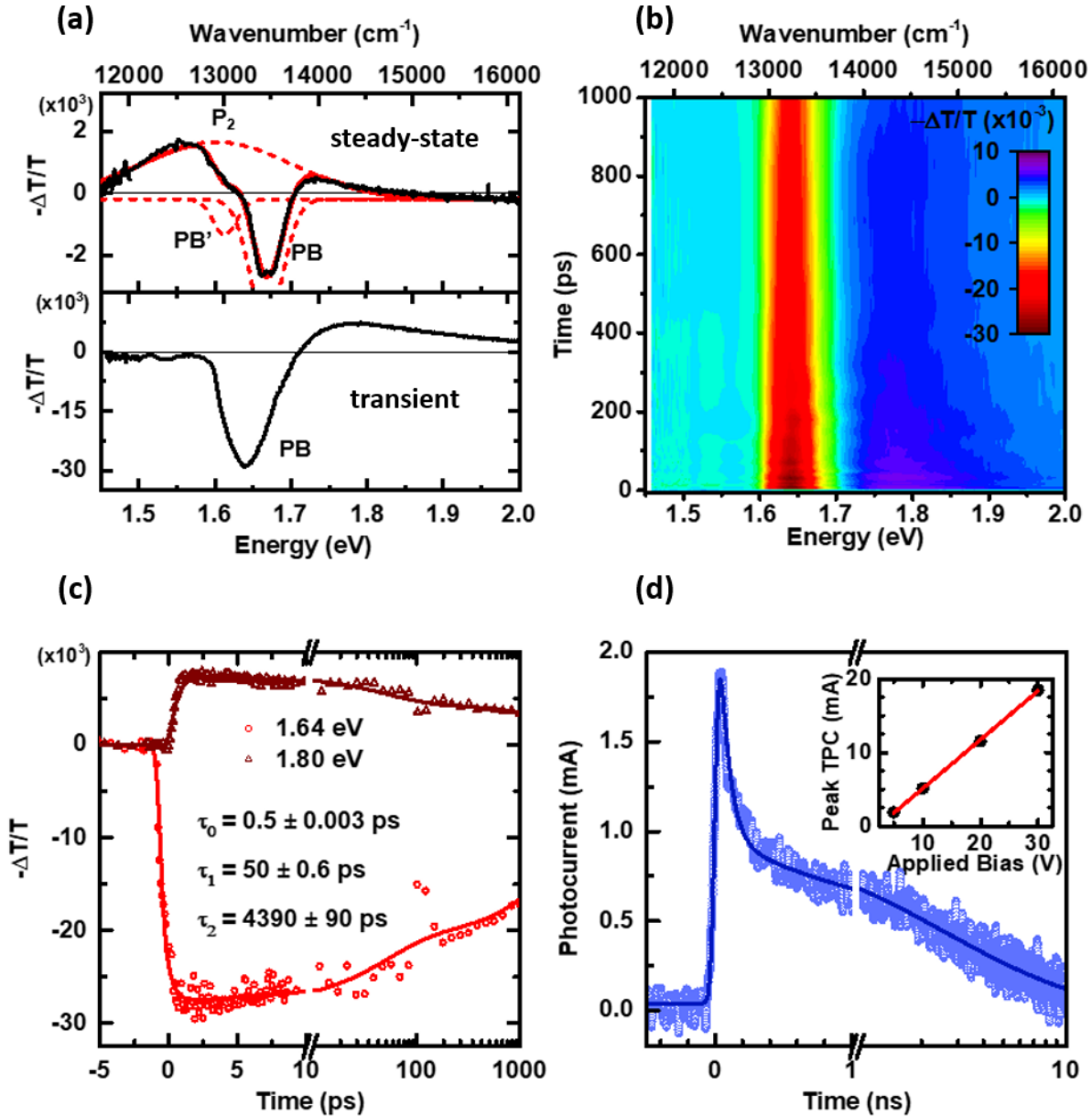

**Figure 5.** Ultrafast polaron formation and decay dynamics. (a) Comparison between steady-state (top panel) and t=1.5 ps transient (bottom panel) photoinduced absorption spectra of MAPbI$_3$. The steady-state spectrum (also shown in Fig. 2) was fitted with three Gaussian curves: P$_2$ centered around 1.60 eV, PB' centered around 1.61 eV, and PB centered around 1.67 eV. (b) 2-dimensional map of transient absorption; global fitting yielded three distinctive lifetimes $\tau_0$=0.500±0.003 ps, $\tau_1$=50.0±0.6 ps and $\tau_2$=4390±90 ps. (c) Temporal slices of transient absorption at 1.64 eV and 1.80 eV, displaying similar decay dynamics. (d) Fast transient photocurrent decay obtained with an applied voltage of 5 V; fitting to a biexponential decay yielded distinctive lifetimes of $\tau_1$ =101±1 ps and $\tau_2$=4220±20 ps. The inset shows the linear dependence of the transient photocurrent peak on applied bias. In (c) and (d), note the shift in timescale from linear to logarithmic in correspondence to the scale break.

All steady-state spectroscopy results seem to point to large polarons being the primary photoexcitations in 3D MAPbI$_3$. To determine their generation and relaxation dynamics, we performed ultrafast (fs) transient absorption and fast (ps) transient photocurrent measurements (Fig. 5). The time evolution of the transient photoinduced absorption (TPA) spectrum upon pulsed laser excitation (t~100 fs, λ= 400 nm, I= 5 nJ/pulse) is shown in the 2D contour plot in Fig. 5b. There are two key features in the TPA spectrum, consisting of a broad absorption peak from 1.7 to 2.0 eV (13700 to 16100 cm$^{-1}$) and a sharp bleaching peak centred at 1.64 eV (13200 cm$^{-1}$), which corresponds to the band gap of MAPbI$_3$. Comparison of the cw-PIA and TPA spectra of MAPbI$_3$ (Figs. 5a, top and bottom panels) shows good qualitative agreement, with coexistence of PIA and PB components at different relative positions. The red-shift of the photobleaching peak in cw-PIA relative to the transient data is likely due to change in bandgap from 1.67 to 1.64 eV from the high temperature tetragonal phase to the low temperature orthorhombic phase,[60] since cw-PIA was collected at ~80 K whereas TPA was obtained at room temperature. Moreover, the TA spectrum contains contributions from both neutral excitons and charged polarons.

The correspondence between early-time TPA and cw-PIA spectra indicates that the long-lived polaronic species are generated at ultrafast time scale (t<100 fs). From the global fitting of the spectral decays (see representative dynamics in Fig. 5c) we determined three distinctive time constants of $\tau_0$= 0.500 ± 0.003 ps, $\tau_1$=50.0 ± 0.6 ps, and $\tau_2$ = 4390 ± 90 ps for sequential exponential decays. We attribute the ultrafast relaxation process of 0.500 ps to lattice thermalization, and the longer time constants to polaron population decay. The recombination process is far to be completed within our experimental temporal range, suggesting a long lived polaron population that is consistent with the observation up to the ms time domain. Note that transient features between 1.7 eV and 2.0 eV have been previously observed in MAPbI$_3$ films,[17-18, 61-62] and assigned to band filling effects by Manser *et al*.,[61] while Zhai *et al*. ascribed them to the photogeneration of free carriers.[62] The long lived decay was also determined in an earlier study, but not attributed to specific features.[17]

The picosecond transient photocurrent (TPC) induced by a femtosecond laser pulse in a photoconductive switch was measured by using a new generation high-speed oscilloscope equipped with a 65 GHz real-time sampling channel (Fig 5d). Unlike all-optical photoinduced absorption measurements that may include signatures of neutral (e.g. excitonic) photogenerated species, fast TPC waveforms provide a direct and selective measurement of charge carrier generation and decay dynamics which correlate directly with transport characteristics of photovoltaic devices. The fast rise time of the TPC signal (limited by the instrument response function) is consistent with the sub-picosecond polaron generation inferred from TPA measurements. The characteristic photocurrent decay times obtained by a biexponential fit with $\tau_1$=101±1 ps and $\tau_2$=4220±20 ps also compare well against the long-lived polaron lifetimes observed in transient absorption ($\tau_1$=50.0 ± 0.6 ps and $\tau_2$=4390±90 ps). Note that the initial dynamics of TPA is likely to include exciton decay, which accounts for its faster early time constant. The inset of Figure 5c shows the linear relation between peak transient photocurrent and applied voltage, which enables the extraction of charge carrier mobility in the linear regime (see Methods). The resulting early-time polaron mobility is of the order of 10 $cm^2V^{-1}s^{-1}$. In comparison, the mobility obtained from steady-state photocurrent measurements of the same sample is of the order of $\mu_{ss} = 0.2 - 2 \text{ cm}^2\text{V}^{-1}\text{s}^{-1}$, assuming a long polaron lifetime extending to the 1-10 ms time scale. Early-time (pre-trapping) mobility values are indeed expected to be larger than steady-state mobility values, which are dominated by thermally activated trapping and detrapping,[63] and closer to band-like mobility estimated by first principle calculations.[3]

Theory for single crystal MAPI yields the mobility values summarized in Table 1 for both positive and negative polarons in the relevant crystallographic directions. The important outcomes are the following:

i) Polaron mobilities are systematically lower than free carrier mobilities, as expected from the enlarged effective masses associated to the phonon clouds. In the low temperature phase, polaron mobilities are 3 to 8 times smaller than free carrier effective masses, while in the high temperature phase between 1.3 to 3 times smaller.

ii) Mobility anisotropy of polarons is smaller than their effective mass anisotropy.

iii) Theory compares reasonably well with data in the literature reporting mobilities of perovskite single crystals from 2.5 to 600 cm$^2$ V$^{-1}$ s$^{-1}$. [49, 64-67]

Our experimental data were obtained in polycrystalline films and cannot be compared directly to single crystal data nor theoretical values. In addition, measurements at different time scales provide different effective mobility values. Just on single crystals[68], reported mobilities vary from almost 1000 to 1 cm$^2$/Vs going from THz (ps time scale) through microwave reflectivity (µs time scale) to space-charge limited current (ms time scale) experimental techniques. In films, mobilities span 2 orders of magnitude, from 0.1 to 80 cm$^2$/Vs. Our results are obtained with complementary techniques not yet reported, and compared reasonably well with the existing literature data.

In conclusion, we have provided a clear indication that the same type of lattice relaxation known to lead to the photogeneration of small polarons in low dimensional perovskites occurs also in the 3D metal halide perovskite MAPbI$_3$. However, in the 3D case, topological constraints limit the extent of lattice deformation due to the balance between the electronic energy and the lattice elastic energy. In this situation, the phonon dressed self-trapped states give rise to large polarons with radius extending over a few lattice sites. Such large polarons are generated at ultrafast time scale and are long lived, with a fraction of the initial population surviving up to the ms time domain, as we see in cw-PIA spectra that clearly indicate vibrational mode frequency renormalization. Large polarons bring about an increase of the effective carrier mass and a concomitant reduction in mobility, overall preserving a large µτ product. Theory accounts for the geometrical relaxation and the vibrational mode renormalization associated to the polaron state, predicts transport anisotropy, and reproduces the low energy polaron electronic spectrum seen in cw-PIA. It also predicts a small charge symmetry breaking, yielding ambipolar transport even in the polaron regime. Carrier recombination is affected by their polaronic nature, for the phonon cloud around the carrier in 3D ionic crystals increases charge screening and reduces Coulomb mediated interactions. This picture validates the idea that large polarons are responsible for the anomalous transport characteristics of metal halide perovskites, including lowering of carrier mobility via rescaling of the effective mass, protection against carrier scattering, and suppression of bimolecular (Langevin) recombination channels.

**Materials and Methods**

**Sample Preparation**

MAPbI$_3$ precursor solutions with concentration 20 wt% and 40 wt% were prepared by dissolving stoichiometric amounts of methylammonium iodide (MAI, Dyesol Inc) and lead iodide (PbI$_2$, 99.99%, TCI) in anhydrous N,N-dimethylformamide (DMF, Sigma-Aldrich). Calcium Fluoride and intrinsic Silicon substrates were sonicated in acetone and isopropyl alcohol (IPA) for 15 minutes followed by 30 minutes of UV/ozone treatment. The substrates were then transferred into a N$_2$ filled glovebox with H$_2$O and O$_2$ levels <1 ppm. The precursor solution was spin-coated on the substrates at 4000 rpm for 30s; toluene dripping was performed after 4s to obtain smooth, high-quality films. The substrates were subsequently annealed at 100º C for 15 minutes. 40 wt% solutions were utilised in PIA measurements to improve the signal to noise ratio while 20 wt% solutions were used for transient absorption and fast photocurrent measurements.

**Steady-State Photoinduced Absorption**

Photoinduced absorption measurements were carried out in a Bruker V80v FTIR spectrometer using a solid-state pump laser (λ=532 nm) with intensity of ~200 mW/cm$^2$. 40 wt% sample was utilized for measurement in the MIR and NIR range while 20 wt% sample was utilized for the visible regime. cw-PIA measurements were carried out at low temperature (78 K) using a Helitran cryostat cooled by liquid nitrogen, pumped at a base pressure of ~5.0×10$^{-5}$ mbar. Four different detectors were used to probe the 3 distinct spectral regimes: Deuterated Triglycine Sulfate (DTGS) and Mercury Cadmium Telluride (MCT) to probe the mid-infrared regime, Indium Gallium Arsenide (InGaAs) to probe near-infrared, and Silicon to probe the visible regime. A notch filter at 532 nm was used for measurements in the visible regime to eliminate pump laser scattering, while a 650 nm band pass filter or a double-polished Silicon wafer were utilized for NIR and MIR measurements, respectively. The FTIR spectrometer operated in rapid scan mode. Transmittance spectra were recorded under photoexcitation, $T_{on}$, and without photoexcitation, $T_{off}$. PIA spectra were then derived as $-\Delta T/T = -(T_{off}-T_{on})/T_{off} \approx \Delta\alpha \cdot d$. where $\Delta\alpha$ is the differential absorption coefficient and $d$ is the film thickness. Total 8000 $T_{on}$ and $T_{off}$ scans each were collected and averaged to obtain the desired signal-to-noise ratio.

**Transient Absorption**

A commercial regenerative amplifier system (Quantronix Integra-C) was used as the laser source at the repetition rate of 1 kHz at 810 nm and pulse width of around 100 fs. A commercial spectrometer, Jobin Yvon CP140-104, equipped with a silicon photodiode array was used to record the transient absorption spectra (Entwicklungsbüro Stresing). A portion of the laser beam was split to generate the 400 nm pump beam using a 1 mm thick BBO crystal cut at 29.2º. White probe light was generated using a sapphire crystal coupled with a 750 nm short pass filter just sufficient to attenuate 800 nm generation beam without saturating the camera and thus, the output spectrum was sensitive from 550 to 830 nm. Another long pass filter with cut-off wavelength of 450 nm was used after the sample to avoid saturation of the camera by the intense pump beam.

Global fitting of the resulting spectral decays was performed using the R-package TIMP on Glotaran interface,[58] with multiple sequential exponential decays. Dispersion compensation was conducted by fitting 2$^{nd}$ order dispersion relationship central at 755 nm in conjunction with the interested multiple exponential decay function. Convergence of the numerical fitting is ensured by multiple rounds of numerical fitting using calculated parameters from previous round, until all parameters stabilize. Overall goodness of the fitting is indicated by the final residual error of 0.489 after stabilization, with no temporal and spectral trend in residue plots.

**Steady-State Photocurrent**

Measurements were performed using conventional amplitude modulation technique in a setup equipped with a Xe lamp source, a monochromator (Horiba iHR550) which disperses light in the wavelength range of 300-900 nm, a mechanical chopper (Stanford SR570), a voltage source (Keithley 6487), and a lock-in amplifier (Stanford SR830). The chopper modulation frequency was 138 Hz, and the lock-in time constant was 300 ms, corresponding to 0.42 Hz equivalent noise bandwidth. Before the measurement, the monochromatic light power/intensity was obtained using a reference calibrated Si photodiode. The responsivity ($R_i$) so obtained is the product of incident power on the sample surface ($P_i$) and measured photocurrent ($iP_h$).

Trap-limited (long-time) charge carrier mobility was estimated using the classical Drude model, according to which the steady-state photocurrent is given by $J_{PC}(t) = e\mu\Delta n(t)F$, where $e$ is the elementary charge, $\mu$ is the charge carrier mobility, $\Delta n$ is the photogenerated carrier density, and $F$ is the applied electric field. The optical generation rate of free carriers can be written as $g(z) = -\eta \frac{1}{\hbar\omega}\frac{dI}{dz} = \eta(1-R)\frac{1}{\hbar\omega}\alpha I_0 e^{-\alpha z}$, where $\alpha$ is the material's absorption coefficient, $R$ its reflectance, $I_0$ the incident illumination intensity, $\eta$ the charge carrier photogeneration quantum efficiency, and $\hbar\omega$ the excitation photon energy. Under steady-state condition, $\Delta n(t) = g\tau = constant$, where $\tau$ is the characteristic carrier lifetime. Thus, assuming unitary quantum efficiency and complete photon absorption, $J_{PC}(t) = e\mu\tau F \int_0^d g(z)\, dz = \frac{I_0 e}{\hbar\omega} F\mu\tau$. As such, with illumination intensity of $I_0$ = 18.5 μW/cm² at photon energy of 1.68 eV, the corresponding mobility-lifetime product is of the order of $\mu\tau = 1.96 \times 10^{-3}\, cm^2 V^{-1}$; assuming a carrier lifetime of $\tau \approx 1 - 10\, ms$ (limited by the cw-PIA modulation frequency), we estimate $\mu_{ss} = 0.2 - 2\, cm^2 V^{-1} s^{-1}$.

**Transient Photocurrent**

The setup utilizes the same femtosecond pump laser as for transient absorption measurements, frequency doubled to the excitation wavelength of 400 nm. Gold contacts were deposited via a shadow mask to form a 50 Ω Auston photoconductive switch, with gap of 0.1 mm and width of 0.6 mm. Measurements were performed in a vacuum chamber focusing 5 mW laser beam power on a 1.5 x $10^{-3}$ $cm^2$ spot size. An external bias of 20 V was supplied by a picoammeter/voltage source (Keithley 6487). The transient photocurrent signal was recorded using a real-time high-speed oscilloscope (Teledyne Lecroy LabMaster 10 Zi-A) equipped with a 65 GHz sampling channel, connected after a Mini-Circuit DC-AC splitter to prevent DC current leakage.

The early-time mobility was estimated from the peak transient photocurrent, $I_{peak} = \eta\phi\mu(1-R)\frac{E_p e}{\hbar\omega}\frac{V}{d^2}$, where $\eta$ is the photogeneration quantum efficiency, $\phi$ is the probability to escape fast recombination, $\mu$ is the charge carrier mobility, $E_p$ is the energy of the laser pulse, $V$ is applied voltage, $d$ is the distance between the electrodes and $\hbar\omega/e$ the photon energy in Volts. We assumed unitary charge generation quantum yield and estimated the probability to escape fast recombination to be 0.1 since the excitation density is on the order of $10^{18}$ $cm^{-3}$ [69]. We estimated reflectance from Fresnel equation, $R = \left|\frac{1-\eta}{1+\eta}\right|^2$ using $\eta = 2.3$ [70], hence $1-R = 0.84$. Using experimental values for the excitation fluence of 0.64 mJ $cm^{-2}$ and d=100 μm, we finally derived $\mu \approx 6$ $cm^2V^{-1}s^{-1}$.

**Computational Methods**

The density functional theory (DFT) calculations were carried out at generalized gradient approximation (GGA)/Perdew-Burke-Ernzerhof (PBE) level using the projector-augmented wave (PAW) method as implemented in the Vienna Ab initio Simulation (VASP) package.[71-72] The plane-wave basis set cutoff of the wavefunctions was set at 400 eV and a uniform grid of 6×6×6 *k*-mesh in the Brillouin zone was employed to optimize the crystal structure of MAPbI$_3$. The van der Waals functional vdW-DF was also included for the structural optimizations and electronic property calculations. The resulting crystal parameter of tetragonal-phase MAPbI$_3$ is *a* = 8.68 Å, *b* = 8.67 Å, *c* = 12.8 Å. A 3×3×3 supercell containing 1296 atoms was used for the large polaron calculations, and the Brillouin zone was sampled by the Γ point. The atomic positions of MAPbI$_3$ supercells in neutral and charged states were fully relaxed until the supercells with forces on each atom less than 0.01 eV/Å. The charge density distribution of valence band maximum (VBM)/conduction band minimum (CBM) for MAPbI$_3$ supercells were used to describe the positive/negative polaron feature. The effective masses for electron ($m_e^*$) and hole ($m_h^*$) were estimated by fitting of the dispersion relation of $m^* = \hbar^2 \left[\frac{\partial^2 \varepsilon(k)}{\partial k^2}\right]^{-1}$ from band structures (Figures S1a and S1b) along the directions Γ-X, Γ-Z and Γ-M for tetragonal phase and Γ-X and Γ-Z for orthorhombic phase. The optical dielectric function was calculated using random phase approximation (RPA) method as implemented in VASP (Figures S1c and S1d).

The infrared and Raman vibrational mode positons and intensities, at the Γ point of the first Brillouin zone, were calculated on orthorhombic-phase MAPbI$_3$ using the Phonon code as implemented in the Quantum Esppresso package.[73] The local density approximation (LDA) exchange-correlation functional with norm-conserving pseudopotentials was used based on the optimized natural and charged structures. The plane-wave expansion cutoff for the wavefunctions was set at 100 Ry. Uniform grids of 12×12×8 Monkhorst-Pack scheme were used for the k-point sampling together with self-consistency threshold of 10$^{-14}$ Ry. The spin-orbit coupling was not included in the Raman calculations since it plays a less significant role than geometry to describe the vibrational properties of heavy-metal based perovskite systems.

**Polaron Mobility**

The electron-phonon coupling is described by the dimensionless Fröhlich parameter $\alpha$, defined as:

$$\alpha = \frac{1}{4\pi\varepsilon_0}\frac{e^2}{\hbar}\left(\frac{1}{\varepsilon_\infty} - \frac{1}{\varepsilon_s}\right)\sqrt{\frac{m}{2\hbar\omega}}$$

where $\varepsilon_0$ is the dielectric constant of vacuum; $\varepsilon_\infty$ and $\varepsilon_s$ are optical and static dielectric constants, respectively; $e$ is the charge of carrier; $2\pi\hbar$ is Planck's constant; $m$ is the bare electron band effective mass; and $\omega$ is the characteristic angular frequency of the longitudinal optical (LO) phonon mode, which is calculated from the Im[$1/\varepsilon(\omega)$] spectra in far-infrared region, as presented in Figure S2.

Ōsaka provided the finite-temperature free energies of the coupling electron-phonon system by extending Feynman's athermal variational solution. The self-free energy of polaron, $F$, under the phono occupation factor $\beta = \omega/k_B T$ was calculated with two parameters $v$ (a unit of $\omega$, the frequency of relation motion between a charge and a coupled LO phonon) and $w$ (a unit of $\omega$).[74-75] The $v$ and $w$ were numerically solved by giving the minimum $F = -(A + B + C)$, where[76]

$$A = \frac{3}{\beta}\left[\ln\left(\frac{v}{w}\right) - \frac{\ln(2\pi\beta)}{2} - \ln\frac{\sinh\left(\frac{v\beta}{2}\right)}{\sinh\left(\frac{w\beta}{2}\right)}\right];$$

$$B = \frac{\alpha v}{\sqrt{\pi[\exp(\beta)-1]}}\int_0^{\frac{\beta}{2}}\frac{\exp(\beta-x)+\exp(x)}{\sqrt{w^2 x\left(1-\frac{x}{\beta}\right)+\frac{Y(x)(v^2-w^2)}{v}}}dx;$$

$$Y(x) = \frac{1}{1-\exp(-v\beta)}[1 + \exp(-v\beta) - \exp(-vx) - \exp\{v(x-\beta)\}];$$

$$C = \frac{3(v^2-w^2)}{4v}\left(\coth\left(\frac{v\beta}{2}\right) - \frac{2}{v\beta}\right);$$

Then the Polaron mobility can be described as:[76-77]

$$\mu = \frac{3\sqrt{\pi}e}{2\pi\omega m\alpha_{e-ph}}\frac{\sinh(\beta/2)}{\beta^{5/2}}\frac{w^3}{v^3}\frac{1}{K};$$

$$\text{where } K = \int_0^\infty \frac{cos(u)}{(u^2+a^2-bcos(vu))^{3/2}} du;$$

$$a^2 = \left(\frac{\beta}{2}\right)^2 + \left(\frac{v^2-w^2}{w^2 v}\right)\beta coth\left(\frac{\beta v}{2}\right);$$

$$b = \left(\frac{v^2-w^2}{w^2 v}\right)\frac{\beta}{sinh\left(\frac{\beta v}{2}\right)};$$

**Free Carrier Mobility**

The free carrier mobilty was calculated by the semi-classical Boltzmann transport theory.[78] Only the contribution of acoustic phonons was considered in evaluating scattering lifetime, where the charge carrier density ($n$) and mobility ($\mu$) are approximated as[79-80]

$$n = \frac{(2m^* k_B T)^{3/2}}{2\pi^2 \hbar^3} \, {}^0F_0^{3/2}; \quad \mu = \frac{2\pi \hbar^4 eB}{m_I^*(2m_b^* k_B T)^{3/2} \Xi^2} \frac{3 \, {}^0F_{-2}^1}{{}^0F_0^{3/2}};$$

$$\text{where } {}^n F_l^m = \int_0^\infty \left(-\frac{\partial f}{\partial \zeta}\right) \zeta^n (\zeta + \alpha \zeta^2)^m [(1 + 2\alpha\zeta)^2 + 2]^{l/2} d\zeta;$$

$$f = 1/(e^{\zeta-\xi}+1); \alpha = k_B T/E_g,$$

$k_B$ is the Boltzmann constant, $e$ is the elementary charge, $T$ is the temperature, $2\pi\hbar$ is the Planck constant, and $\xi$ is the reduced chemical potential; $m^*$ is the density of state effective mass, $m_I^*$ is the conductivity effective mass, $m_b^*$ is the band effective mass; $B$ is the bulk modulus ($B = \partial^2 E/\partial V^2$), $\Xi_{e-p/h-p}$ is the electron-phonon (or hole-phonon) coupling energy ($\Xi_{e-p/h-p} = V_0(\Delta E_{CBM/VBM}/\Delta V)$, $n$, $m$, and $l$ power integer indices, $E_g$ is the electronic band gap, and $\zeta$ is the reduced carrier energy.

**Acknowledgements**

We would like to thank Dr. Zilong Wang for the preliminary transient photocurrent data and assistance from Teledyne LeCroy Application Engineer Mr. Wayne Lim. W.P.D.W. is grateful to Dr. Meng Lee Leek and Dr. Vijila Chellappan for fruitful discussions. Research was supported by the National Research Foundation (NRF-CRP14-2014-03) and by the Ministry of Education (MOE2016-T1-1-164) of Singapore. Work at King Abdullah University of Science and Technology was supported by the KAUST Supercomputing Laboratory.